\documentclass[twocolumn,superscriptaddress]{revtex4}

\usepackage{graphicx}

\newcommand{\up}{\uparrow}
\newcommand{\dn}{\downarrow}
\newcommand{\dg}{^\dagger}
\newcommand{\bsigma}{{\mbox{\boldmath $\sigma$}}}
\newcommand{\beps}{{\mbox{\boldmath $\epsilon$}}}

\begin{document}

\title{Ultra-long distance interaction between spin qubits}

\author{Guido Burkard}

\affiliation{Department of Physics and Astronomy, University
of Basel, Klingelbergstrasse 82, CH-4056 Basel, Switzerland}

\author{Atac Imamoglu}

\affiliation{Institute of Quantum Electronics, ETH-Z\"{u}rich, CH-8093,
Z\"{u}rich, Switzerland}


\begin{abstract}
We describe a method for implementing deterministic quantum gates
between two spin qubits separated by centimeters.  Qubits defined
by the singlet and triplet states of two exchange coupled quantum
dots have recently been shown to possess long coherence times.
When the effective nuclear fields in the two asymmetric quantum
dots are different, total spin will no longer be a good quantum
number and there will be a large electric dipole coupling between
the two qubit states: we show that when such a double quantum dot
qubit is embedded in a superconducting microstrip cavity, then
the strong coupling regime of cavity quantum electrodynamics lies
within reach. Virtual photons in a common cavity mode could
mediate coherent interactions between two distant qubits embedded
in the same structure; the range of this two-qubit interaction is
determined by the wavelength of the microwave transition.
\end{abstract}

\maketitle

\textit{Introduction.} Experimental realization of conditional
quantum dynamics of two isolated solid-state quantum systems has
become a holy grail of mesoscopic physics research, due to its
potential implications for scalable quantum information
processing.  The majority of theoretical proposals aimed at this
goal is based on nearest-neighbor interactions such as the
Heisenberg exchange coupling between quantum dot (QD) spins
\cite{LD}. However, to achieve lower accuracy thresholds
for quantum error correction, the
implementation of coherent long-range interactions between two
qubits is highly desirable \cite{Imamoglu99}. 
Optical dipole-dipole interactions \cite{Calarco2003},
capacitive coupling \cite{Taylor2005}
and optical cavity-mediated interactions \cite{Imamoglu99} 
between spins could be used to realize
controlled quantum gate operations on length-scales comparable to
optical wavelengths: these mechanisms may then enable coherent
interactions between a limited number ($\le 10$) of QD spins.

In this Rapid Communication, we show how ultra-long range coherent interactions
between two spin qubits separated by centimeters can be
mediated. Our proposal is motivated by two recent remarkable
experimental achievements: 1) realization of circuit-QED using a
Josephson-junction charge qubit strongly coupled to a
superconducting (SC) microstrip cavity \cite{Wallraff};
and 2) demonstration that the
singlet-triplet subspace of a double-QD structure constitutes a
promising qubit exhibiting coherence times exceeding $10\,\mu{\rm s}$
\cite{Petta,Koppens}.
We show here that due to the presence of magnetic field gradients
caused by partially polarized QD nuclear spin ensembles, it is possible to
induce a large electric-dipole coupling between singlet (S) and
triplet ($T_0$) states by adjusting an external gate voltage \cite{Lai2006}. 
Since the energy of the S-$T_0$ transition is in the microwave range, it is
possible to use SC microstrip cavities with a length
($L$) equal to the transition wavelength ($\lambda$) and a
cavity-volume $V_{\rm cav} \sim 10^{-8} \lambda^3$ to mediate
interactions between two qubits embedded in the same cavity via
virtual microwave photon exchange. A distinguishing feature of our
proposal is the large separation between the length scales
determining single qubit control, determined either by fabrication
($\sim 50\,{\rm nm}$) or optical wavelength ($\sim \,1 \mu{\rm m}$),
and that of
two-qubit interactions, ultimately determined by the wavelength
corresponding to the (adjustable) S-$T_0$ qubit transition.
As in Ref.~\onlinecite{Taylor2005}, the qubits here are coupled 
via their electric dipole moment:  however, the use of a cavity
does away with the $1/r^3$ decay of the dipolar interaction that
limits the range of the coupling in Ref.~\onlinecite{Taylor2005}.
In contrast to the scheme introduced in Ref.~\onlinecite{Childress2004}
which couples spins via their \textit{magnetic} dipole moment,
our proposal does not require the use of electron spin resonance (ESR).

\begin{figure}
\centerline{\includegraphics[width=9cm]{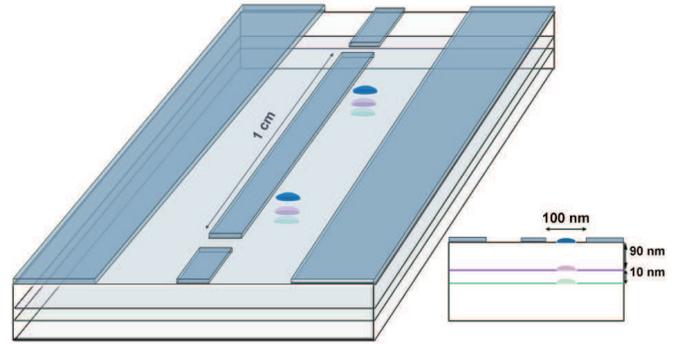}}
\caption{
(Color online) The proposed setup with two vertically coupled double QDs next to a
superconducting micro-strip cavity.
The topmost QD serves as a marker.
\label{fig:setup}}
\end{figure}
Fig.~\ref{fig:setup} shows the structure that we envision: the
micro\-strip cavity is defined by a wavelength-long center
SC strip separated from the ground planes by $\sim 100\,{\rm
nm}$. The whole structure is deposited on a Molecular Beam Epitaxy
grown GaAs wafer containing a stack of two self-assembled QDs that
are tunnel coupled and buried $\sim 100\,{\rm nm}$ below the
surface. Finally, $\sim 50\,{\rm nm}$ below the lowest QD layer
lies either an n-doped $20\,{\rm nm}$ GaAs layer or a
modulation-doped quantum well. The ohmic contact to this bottom
electron reservoir allows for applying a gate voltage $V_{\rm
gate}$ that is used to inject single electrons into each QD
deterministically and to bring the electronic states of the two
QDs in and out of resonance \cite{footnote1}. Even though the
self-assembled QDs in the first layer nucleate at random
locations, the QDs of the second layer have a very high likelihood
for nucleating directly above the QDs of the first layer. It has
been shown that Atomic Force Microscopy can be used to
determine the position of stacked QDs with a spatial resolution of
$25\,{\rm nm}$ \cite{AFM}. 
In order to apply independent gate
voltages to two double QDs embedded in the same cavity, it will be
necessary to wafer-fuse two separate samples before depositing the
SC thin layers \cite{footnote2}.

\textit{Coupling mechanism.} Coupling the spin singlet $|S\rangle$
and triplet $|T_0\rangle$ in a double QD (the two qubit states)
via the emission or absorption of a cavity photon requires a
sufficiently strong electric dipole transition between the two
states \cite{footnote3}. The key question is under what
conditions it is possible to obtain an electric dipole transition
in a double QD. First, we remark that the two QDs need to be
coupled via inter-dot tunneling (with a tunneling energy $t$) for
a non-zero dipole matrix element; indeed, we find below in
Eq.~(\ref{g-off}) that the matrix element is proportional to the
singlet-triplet energy splitting, i.e. the exchange energy
$J\propto t^2$ \cite{BLD}.

Provided that $|t|>0$, there are still two
independent symmetries that can prevent electrical dipole transitions.
One of the two symmetries derives from the spin-conserving nature
of the electron-photon interaction. The spin singlet $|S\rangle$
and triplets $|T_{0,\pm}\rangle$ are eigenstates of the total spin
with different spin quantum numbers $S=0$ and $S=1$ and cannot be
transformed into each other by the emission or absorption of a
photon which changes only the orbital angular momentum. However,
$|S\rangle$ and $|T_{0}\rangle$ are mixed, and thus the spin
selection rules broken, by the presence of a magnetic field that
is inhomogeneous on the scale of the inter-dot distance. Such a
field inhomogeneity $\delta h$ is usually unavoidable in the form
the Overhauser field due to the hyperfine coupling of the electron
spin to the surrounding nuclear spins in the QD material. The
dipole matrix element given below in Eq.~(\ref{g-off}) is indeed
proportional to $\delta h$.

\begin{figure}
\centerline{\includegraphics[width=8cm]{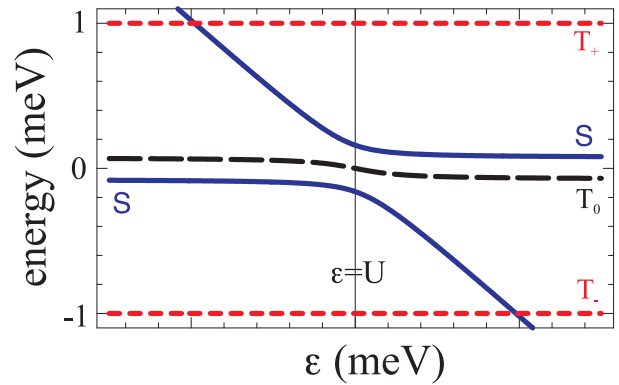}}
\caption{
(Color online) Energy of the two-electron states in a double QD as a function
of the inter-dot detuning $\varepsilon$ around the resonance
$\varepsilon=U$, indicated by a dotted vertical line.
The parameters chosen for the plot are $U=10\,{\rm meV}$, $t=0.1\,{\rm meV}$,
$\delta h=0.15\,{\rm meV}$, and $g\mu_B B=1\,{\rm meV}$.
\label{fig:levels}}
\end{figure}

The second problem to be overcome if dipole transitions are
to occur between $|S\rangle$ and $|T_0\rangle$ in a double QD is the
orbital symmetry that exchanges the two QDs.  The effect
of tunneling on $|S\rangle$
consists in the admixture of the $|S(1,1)\rangle$ singlet
(one electron in each QD) with the states $|S(2,0)\rangle$
and $|S(0,2)\rangle$ that
involve two electrons on the same QD;  for symmetric QDs,
this admixture is restricted to the symmetric combination
$|D_+\rangle=(|S(2,0)\rangle+|S(0,2)\rangle)/\sqrt{2}$ of
doubly occupied states on the two QDs,
while the electric dipole Hamiltonian has odd parity
(it can be represented in terms of the momentum or the
position operator, both having odd parity) and thus
couples the singlet (even parity) exclusively
to the antisymmetric combination
$|D_-\rangle=(|S(2,0)\rangle-|S(0,2)\rangle)/\sqrt{2}$,
and thus not to the "qubit" singlet.
The mirror symmetry between the QDs is broken if the QDs are
electrically biased, thus detuning their single-electron levels by
an energy $\varepsilon$. 
Roughly speaking, the electric bias $\varepsilon$ creates a situation with a
mobile charge that allows for an electric dipole moment which is absent in the
unbiased double QD ($\varepsilon=0$).
We plot the two-electron spectrum in a
double QD as a function of $\varepsilon$ in Fig.~\ref{fig:levels}.
We expect that the dipole matrix element is
proportional to $\varepsilon$.

Having discussed the underlying physical considerations, we
proceed with the key results of the paper and defer the derivation
of qubit-cavity coupling strength $g$ to the last part of the
Letter. In the presence of electric (or magnetic) dipole coupling
between the states  $|S\rangle$ and $|T_0\rangle$ that define our
qubit, the Hamiltonian is
\begin{equation}
  \label{qubit-cavity}
  H = \frac{\bar\varepsilon}{2}\sigma_z + g\sigma_x \left(a+a^\dagger\right),
\end{equation}
with $\bar\varepsilon$ the $S$-$T_0$ splitting and,
for electric dipole coupling,
\begin{equation}
  \label{g-off}
  g =  e a E_0 \frac{J}{\hbar\omega}
      \frac{\varepsilon \, (\delta h /2) }{U^2 -\varepsilon^2 -(\delta
      h/2)^2}.
\end{equation}
The vacuum value of the electric field is given by $E_0 =
V_{\rm rms}^0/d = \sqrt{\hbar\omega/2\epsilon_0\epsilon L d^2}$
\cite{Blais2004}, where $L$ is the length of the center
SC, $d$ is its separation from the ground
SC planes, and $\epsilon$ is the effective dielectric
constant seen by the cavity mode. The operators $a^\dagger$ and
$a$ create and annihilate a cavity photon with frequency
$\omega/2\pi$. The energy denominator in Eq.~(\ref{g-off}) arises
from the admixture of the $S(1,1)$ and $T_0$ states with the
doubly occupied states $S(2,0)$ and $S(0,2)$ that are separated in
energy by $U\pm\varepsilon \pm\delta h/2$. The details of the
derivation of Eqs.~(\ref{qubit-cavity}) and (\ref{g-off}) will be
given further below.

\textit{Two-qubit coupling.} We now turn to the situation of two
double QDs coupled to the same cavity, as shown in
Fig.~\ref{fig:setup}. By introducing the rotating wave approximation
in Eq.~(\ref{qubit-cavity}) and eliminating the cavity mode
using a Schrieffer-Wolff (SW) transformation
\cite{Madelung,Imamoglu99}, we obtain
\begin{equation}
  \label{eq:3}
  H_{\rm eff} = \sum_{i=1,2} \frac{\tilde\varepsilon_i}{2}\sigma_z^{(i)}
                + g_{\rm eff} \left(\sigma_+^{(1)}\sigma_-^{(2)}+\sigma_-^{(1)}\sigma_+^{(2)}\right),
\end{equation}
with the effective qubit-qubit coupling parameter
$g_{\rm eff} = g_1 g_2 [1/(\bar\varepsilon_1-\hbar \omega)
+1/(\bar\varepsilon_2-\hbar \omega)]$,
and the Stark-shifted single-qubit splitting,
$\tilde \varepsilon_i /2
= \bar \varepsilon_i /2 +g_i^2(\langle n\rangle+1/2)
/(\bar\varepsilon_i-\hbar\omega)$,
where $\langle n\rangle=\langle a^\dagger a\rangle$ denotes the
number of photons in the cavity.

\textit{Qubit-cavity coupling.}
To derive Eqs.~(\ref{qubit-cavity}) and (\ref{g-off}), we consider
the Hamiltonian of a single qubit in the cavity,
\begin{equation}
\label{H}
H=H_{\rm el} + H_{\rm cav} + H_{\rm dip},
\end{equation}
describing the electronic degrees of freedom, the cavity field
$H_{\rm cav} = \hbar\omega (a^\dagger a+1/2)$, and the electric dipole
coupling between the qubit and the cavity. In the first step of
our derivation, we concentrate on the electronic part: we write
$H_{\rm el} = H_D + H_T + H_{\rm int}$ where
\begin{equation}
  \label{HD}
  H_D = \sum_{\stackrel{\alpha=L,R}{\sigma,\sigma'=\up,\dn}}
              c_{\alpha\sigma'}\dg \left(\varepsilon_{\alpha}
                 + \frac{\hbar}{2}g\mu_B {\bf B}_\alpha \cdot \bsigma_{\sigma'\sigma}\right)c_{\alpha\sigma},
\end{equation}
denotes the single-electron Hamiltonian of the
lowest-energy orbital on each separate QD,
$H_T = t \sum_{\sigma=\up,\dn}\left(c_{L\sigma}\dg c_{R\sigma} + c_{R\sigma}\dg c_{L\sigma}\right)$
accounts for electron tunneling between the QDs, and
$H_{\rm int}=U\sum_{\alpha=L,R} c_{\alpha\up}\dg c_{\alpha\up}c_{\alpha\dn}\dg c_{\alpha\dn}$
describes the Coulomb interaction between two electrons occupying
the same QD. In Eq.~(\ref{HD}), $\varepsilon =
\varepsilon_{L}-\varepsilon_R$ is the asymmetry of the double QD
and ${\bf B}_{L,R} = {\bf B}\pm \delta{\bf B}/2$ the effective
magnetic field for an electron on the left (right) QD. The
presence of nuclear spins in the QDs gives rise to inhomogeneous
effective magnetic (Overhauser) fields ${\bf B}_{L,R} = \sum_i A_i
{\bf I}_{L,R}^i$, where ${\bf I}_{L,R}^i$ is the $i$th nuclear
spin in the QD $L,R$, and $A_i$ is the corresponding hyperfine
coupling constant \cite{footnote4}. The operators  $c_{\alpha\sigma}^\dagger$
($c_{\alpha\sigma}$) create (annihilate) an electron in an
orthogonalized Wannier orbital $\Phi_{L,R} = (\varphi_{L,R}-
\gamma \varphi_{R,L})/\sqrt{1-2 S \gamma+\gamma^2}$ in the QD
$\alpha=L,R$ with spin $\sigma=\uparrow,\downarrow$, where
$S=\langle \varphi_L|\varphi_R\rangle$ denotes the overlap
integral between the left and right unnormalized orbitals and
$\gamma=(1-\sqrt{1-S^2})/S$. In parabolic QDs, the wavefunctions
$\varphi_\alpha$ are Gaussian.

A low energy two-electron double QD where only the ground orbital
state on each QD can be occupied, has six possible states: the
three spin triplets $|T_0\rangle  = \frac{1}{\sqrt{2}}\left(
c_{L\up}\dg c_{R\dn}\dg + c_{L\dn}\dg
c_{R\up}\dg\right)|0\rangle$, $|T_\sigma\rangle  = c_{L\sigma}\dg
c_{R\sigma}\dg|0\rangle$ ($\sigma=\up,\dn$) with $S_z=0,\pm 1$,
and the three spin singlets $|S\rangle  \equiv |S(1,1)\rangle =
\frac{1}{\sqrt{2}}\left( c_{L\up}\dg c_{R\dn}\dg - c_{L\dn}\dg
c_{R\up}\dg\right)|0\rangle$, and $|D_\pm\rangle =
\frac{1}{\sqrt{2}}\left(c_{L\up}\dg c_{L\dn}\dg \pm c_{R\up}\dg
c_{R\dn}\dg\right)|0\rangle$, all with $S=S_z=0$. Here, $D_\pm$
are linear combinations of the states with double occupation of a
QD and $|0\rangle$ is the state with no electrons.

We choose a coordinate system such that the $z$ axis is along the
homogeneous part of the field ${\bf B}$ and decompose the
difference field into its longitudinal and transverse parts,
$\delta {\bf B}=\delta {\bf B}_z + \delta {\bf B}_\perp$. We
assume $\delta B_{\perp} \ll B_z$ which ensures that
the spin-polarized states $|T_\sigma\rangle$ are decoupled from
the remaining four states \cite{footnote5}.  We can then write the
Hamiltonian as the four-by-four matrix in the basis spanned by
$|T_0\rangle$, $|S\rangle$, $|D_+\rangle$, and $|D_-\rangle$,
\begin{equation}
  \label{H4}
  H = \left(\begin{array}{c c c c}
0                  & \delta h/2    & 0  & 0 \\
\delta h/2         & 0             & 2t & 0 \\
0                  & 2t            & U  & \varepsilon \\
0                  & 0             & \varepsilon  & U
\end{array}\right),
\end{equation}
where we have introduced the relative Zeeman energy
$\delta h=g\mu_B \delta B_z$ between the dots.
For $\varepsilon=0$, the
$|D_-\rangle$ state completely decouples because its orbital
symmetry forbids any coupling to the other singlets, while a
coupling to the triplet is impossible due to spin conservation.

In the weak tunneling regime $t\ll U-\varepsilon$ we can eliminate
$|D_\pm\rangle$ by means of a SW transformation \cite{Madelung}
$\tilde{H} = e^{-S} H e^S \simeq H_0 + \left[H_T,S\right]/2$,
with $S=-S^\dagger$ and $H_0=H_D+H_{\rm int}$.  
The terms of order $H_T$ are cancelled in $\tilde{H}$
because we choose $S$ such that $[H_0,S] = -H_T$,
\begin{equation}
  \label{S}
  S = \frac{4t}{(U^2-\varepsilon^2)^2-2\delta h^2(U^2+\varepsilon^2)}\left(\begin{array}{c c}
0  & s\\
-s & 0
\end{array}\right),
\end{equation}
with
\begin{equation}
  \label{s}
  s = \left(\begin{array}{c c}
      \delta h (U^2+\varepsilon^2-(\delta h/2)^2) & -2\delta h U\varepsilon\\
      2 U (U^2-\varepsilon^2-(\delta h/2)^2) & -2\varepsilon (U^2-\varepsilon^2+(\delta h/2)^2)
\end{array}\right).
\end{equation}
We assume here that we are in the regime
$U\gg t,\delta h$ and allow $\varepsilon$ to lie in the whole range
$0\le \varepsilon\le U$.
The effect of the SW transformation is to separate the
states with single occupation from $|D_\pm\rangle$
within the lowest order in $t/(U-\varepsilon)$ in the Hamiltonian,
\begin{equation}
\label{Htilde}
\tilde{H} \simeq \left(\begin{array}{c c}
\tilde{H}_S & 0\\
0           & \tilde{H}_D
\end{array}\right),
\quad
  \tilde{H}_S \simeq \left(\begin{array}{c c}
      0 & \delta \tilde h/2  \\
      \delta \tilde h/2 & -J
\end{array}\right),
\end{equation}
with the exchange coupling
\begin{equation}
  \label{J}
  J = \frac{4 t^2 U (U^2-\varepsilon^2-(\delta h/2)^2)}{(U^2-\varepsilon^2)^2-2(\delta h/2)^2(U^2+\varepsilon^2) +(\delta h/2)^4},
\end{equation}
and the effective relative field
\begin{equation}
  \label{dhtilde}
  \delta\tilde h = \delta h\left(1 - \frac{J(U^2+\varepsilon^2-(\delta h/2)^2)}{4U(U^2-\varepsilon^2-(\delta h/2)^2)}\right).
\end{equation}
Note that for $\varepsilon=\delta h=0$, Eq.~(\ref{J}) reduces to
the familiar expression $J=4t^2/U$. From Eq.~(\ref{Htilde}), we
obtain the singlet and triplet eigenenergies
\begin{equation}
  \label{eigenenergies}
  \bar\varepsilon_\pm = \frac{1}{2}\left(-J \pm \sqrt{J^2 + \delta \tilde h^2}\right)
                   \equiv -\frac{J}{2} \pm \frac{\bar\varepsilon}{2} .
\end{equation}

\textit{Dipole matrix element.} Optical transitions conserve spin,
therefore transitions between the singlets $|S\rangle$,
$|D_\pm\rangle$ and the triplets $|T_i\rangle$ ($i=0,\up,\dn$) are
forbidden.  However, the presence of an inhomogeneous magnetic
field $\delta {\bf B}$ breaks spin
symmetry and allows for electric dipole transitions.
The dipole coupling to a single cavity mode is described by \cite{CT}
\begin{equation}
  \label{H_dip}
  H_{\rm dip} = -\frac{e}{m}{\bf A}_0\cdot {\bf p}
      = -\frac{e}{m}\left(\frac{\hbar}{2\epsilon_0\epsilon V\omega}\right)^{1/2}\beps\cdot{\bf p}\left(a + a^\dagger\right),
\end{equation}
where $V = L d^2$, ${\bf A}_0$ denotes the vector potential at
${\bf r}=0$ and $a^\dagger$ ($a$) creates (annihilates) a cavity
photon with frequency $\omega$, described by $H_{\rm cav}$.
In the following, we determine the dipole matrix element
\begin{equation}
  \label{g}
  g = -\frac{e}{m}\left(\frac{\hbar}{2\varepsilon_0\varepsilon V\omega}\right)^{1/2}
          \langle \bar T_0|\beps \cdot {\bf p}| \bar S \rangle,
\end{equation}
where $| \bar S \rangle$ and $| \bar T_0 \rangle$ are the eigenstates
of $\tilde H_S$ in Eq.~(\ref{Htilde}).

In the single-particle eigenbasis of $H_D+H_T$, i.e.,
the bonding and antibonding orbitals
$|\Phi_{\pm,\sigma}\rangle
= (c^\dagger_{L\sigma}\pm c^\dagger_{R\sigma})|0\rangle/\sqrt{2}$,
and using
$\langle n|{\bf p}|m\rangle
= -im\langle n|[H_D+H_T,{\bf x}]|m\rangle/\hbar$,
we find
$\langle\Phi_\pm |{\bf p}|\Phi_\pm\rangle = 0$
and
$\langle\Phi_- |{\bf p}|\Phi_+\rangle = (-imt/\hbar)
         [ \langle\varphi_L|{\bf x}|\varphi_L\rangle
               -\langle\varphi_R|{\bf x}|\varphi_R\rangle
 + 2i {\rm Im}\langle \varphi_L|{\bf x}|\varphi_R\rangle]
/2\sqrt{1-S^2}$.
The last term can have a nonzero component
perpendicular to ${\bf x}$ due to orbital diamagnetism,
but it turns out that this real contribution to the single-particle
matrix element does not contribute to $g$.
We find $\langle\Phi_- |p_x|\Phi_+\rangle = imta/\hbar$,
while the other components have vanishing imaginary parts.
The electric dipole moment of the double QD is thus directed
vertically in Fig.~\ref{fig:setup} and couples to the 
vertical component of the cavity field which is
increased by positioning the QDs close to the microstrip.

The only nonvanishing two-electron matrix element between the unperturbed
states $|S\rangle$, $|T_0\rangle$, and $|D_\pm\rangle$ is
\begin{equation}
  \label{dipole-4}
  \langle D_-|p_x|S\rangle = 2i\, {\rm Im}\langle \Phi_-|p_x|\Phi_+\rangle
                               = 2imta/\hbar .
\end{equation}
We can transform $H_{\rm dip}$ into the new basis using the same
SW transformation $\tilde H_{\rm dip} \simeq H_{\rm dip}
  +\left[H_{\rm dip},S\right]$. In the subspace spanned by the transformed states
$|\tilde S\rangle$ and $|\tilde T_0\rangle$,
we obtain for the momentum operator in $H_{\rm dip}$
\begin{equation}
  \label{dipole-5}
  \tilde p_x = \frac{i a m}{\hbar}\frac{\varepsilon J \, \delta h/2}{U^2 -\varepsilon^2 -(\delta h/2)^2 }
\left(\begin{array}{c c}
      0 & -1\\
      1 & 0
\end{array}\right).
\end{equation}
Transforming Eq.~(\ref{dipole-5})
into the eigenbasis of Eq.~(\ref{Htilde}) is a rotation in the
(pseudo)-xz plane thus leaving the dipole Hamiltonian Eq.~(\ref{dipole-5}),
having the form of a pseudo-field in y-direction, invariant up to
a phase factor.
With Eq.~(\ref{g}), the qubit-cavity Hamiltonian Eq.~(\ref{H}) in
the logical subspace of the singlet-triplet qubit takes the form
of Eq.~(\ref{qubit-cavity}) with the coupling constant
Eq.~(\ref{g-off}). Close to resonance, we can replace
$\hbar\omega$ by $\bar\varepsilon = \bar\varepsilon_+  -
\bar\varepsilon_- =\sqrt{J^2+\delta \bar h^2}$ in
Eq.~(\ref{g-off}). For $\hbar \omega\approx 0.1\,{\rm meV}$,
$\epsilon\simeq 13$ (GaAs), $V=1\,{\rm cm}(100\,{\rm nm})^2$, we
arrive at a vacuum field of $E_0\approx 25\,{\rm V/m}$.  For
self-assembled QDs, we estimate the dot-distance to be of the
order of $a \approx 10\,{\rm nm}$, therefore $ea E_0/h \simeq
0.25\,\mu{\rm eV} \simeq 65\,{\rm MHz}$.

For further discussion, we introduce $\Delta^2 = U^2 -
\varepsilon^2$ and work in the regime of a strongly biased QD
pair, $\Delta\ll U$. We can envision a hierarchy of energy
scales $\delta h, J\ll \Delta \ll U$, e.g., $J\approx \delta
h\approx 0.1\,{\rm meV}$ (for recent experimental results on
optically generated nuclear polarization in QDs, 
see Ref.~\onlinecite{Lai2006}),
$\Delta\gtrsim 1\,{\rm meV}$, and
$U\approx 10\, {\rm meV}$. In this case $\delta \bar h\approx
\delta h$, and near resonance $g \approx ea E_0$. Thus, 
$g \gg \omega/Q, \gamma$, provided that the cavity quality factor $Q
> 10^4$ and the spin decoherence rate $\gamma < 10^7\,{\rm s}^{-1}$. 
The SW transformation can be applied if $t/\sqrt{\Delta^2-\delta h^2}\ll
1$:  hence the ``resonance'' $\Delta\rightarrow\delta h$ where
formally $g\rightarrow eaE_0$ is not within the regime of the
validity of our result. If $J\ll \delta h\ll \Delta\ll U$, we can
simultaneously satisfy $U\delta h/\Delta \approx 1$ ($g\approx
eaE_0$) and $t/\Delta\ll 1$. In the regime $\varepsilon, \delta
h\ll U$,  we obtain $\delta\bar h\approx \delta h$ and therefore
$g\simeq ea E_0 \varepsilon\, \delta h/U^2 \ll ea E_0$.

\textit{Conclusions.} A cavity-double-QD coupling strength of $g
\sim 65 \,{\rm MHz}$ implies that it is possible to implement
two-qubit gates on timescales $\sim 10\,{\rm ns}$: while this is
already three orders of magnitude shorter than the spin (memory)
decoherence time, an important open question that needs to be
addressed is the gate errors. The fact that total spin is not a
good quantum number during the gate operation when the
dipole-coupling of $|S\rangle$ and $|T_0\rangle$ states is on,
most likely introduces additional phonon or charge fluctuation
mediated decoherence.   However, even in the case of relatively
strong charge decoherence, one could still envision using the
cavity-mediated coupling as a source of entanglement that can be
distilled and used for remote gate operations. We also emphasize
that the observation of conditional quantum dynamics between two
spins with a macroscopic separation would itself be an exciting
goal for the emerging field of spintronics.

We acknowledge discussions with A. Wallraff, A. Badolato,
M. Stiller, W. Coish, M. Trif, and D. Loss.  This research is
supported by NCCR-Nanoscience and SNF.  After completion of this
manuscript, we became aware of a related manuscript under
preparation \cite{TaylorLukin}.


\begin{thebibliography}{99}
\bibitem{footnote1}
The exact distance of the
electron reservoir from the QDs and its electron density can be
adjusted to maximize the cavity quality-factor.
\bibitem{footnote2}
We assume that $Q > 10^4$ can be obtained for microstrip
cavities fabricated on GaAs wafer that contains a narrow doped
layer $200\,{\rm nm}$ below the surface.
\bibitem{footnote5}
The analysis including the
polarized triplet states $|T_\sigma\rangle$ is straight-forward
but does not lead to additional dipole transitions within the
low-energy subspace (singly occupied QDs).
\end{thebibliography}
\end{document}